%% file: ifacconf.tex
\documentclass{ifacconf}

\usepackage{graphicx}      
\usepackage{natbib}        

\usepackage{amsmath}
\usepackage{amsfonts}
\usepackage{cases}
\usepackage{caption}
\usepackage{subcaption}
\usepackage{booktabs}
\usepackage{multirow}
\usepackage[table]{xcolor}

\newtheorem{theorem}{Theorem}

\theoremstyle{definition}
\newtheorem{definition}{Definition}

\usepackage[acronym]{glossaries}
\newacronym{CPHSs}{CPHS}{Cyber-Physical-Human Systems}
\newacronym{EV}{EV}{Electric Vehicle}
\newacronym{DPG}{DPG}{Dynamic Population Game}
\newacronym{DPGs}{DPGs}{Dynamic Population Games}
\newacronym{SNE}{SNE}{Stationary Nash Equilibrium}
\newacronym{MDP}{MDP}{Markov Decision Process}
\newacronym{iid}{i.i.d.}{independent and identically distributed}

\newcommand{\Nat}{{\mathbb N}}
\newcommand{\Real}{{\mathbb R}}
\newcommand{\R}{{\mathcal R}}
\newcommand{\T}{{\mathcal T}}
\newcommand{\U}{{\mathcal U}}
\newcommand{\K}{{\mathcal K}}
\newcommand{\X}{{\mathcal X}}
\newcommand{\D}{{\mathcal D}}
\newcommand{\B}{{\mathcal B}}
\newcommand{\Out}{{\mathcal O}}
\newcommand{\highway}{{\texttt H}}
\newcommand{\parking}{{\texttt P}}
\newcommand{\pr}{{\textup{pr}}}
\newcommand{\gp}{{\textup{gp}}}
\newcommand{\spr}{{s^\pr}}
\newcommand{\sgp}{{s^\gp}}
\newcommand{\suburb}{{\texttt S}}
\newcommand{\city}{{\texttt C}}
\newcommand{\Prob}{{\mathbb P}}
\newcommand{\dnt}{{\neg}}
\newcommand{\bmax}{{b^\textup{max}}}
\newcommand{\kbar}{{\bar k}}
\newcommand{\kmax}{{k^\textup{max}}}
\newcommand{\tqueue}{{t^\textup d}}
\newcommand{\sigmanom}{\sigma^\textup{n}}
\newcommand{\paybar}{\bar p}
\newcommand{\all}{\textup{all}}
\newcommand{\act}{\textup{act}}
\newcommand{\paytilde}{\tilde p}
\newcommand{\sigmabar}{\bar \sigma}
\newcommand{\en}{\textup{en}}
\newcommand{\ex}{\textup{ex}}
\newcommand{\sigmabarendo}{\sigmabar^\en}
\newcommand{\sigmabarexo}{\sigmabar^\ex}
\newcommand{\sw}{\textup{SW}}
\newcommand{\swendo}{\sw^\en}
\newcommand{\swexo}{\sw^\ex}
\newcommand{\bench}{\textup{ben}}
\newcommand{\sigmabarbench}{\sigmabar^\bench}

\newcommand{\floor}[1]{\left\lfloor #1 \right\rfloor}
\newcommand{\ceil}[1]{\left\lceil #1 \right\rceil}
\DeclareMathOperator*{\argmax}{arg\,max}
\begin{document}
\begin{frontmatter}

\title{To Travel Quickly or to Park Conveniently: Coupled Resource Allocations with Multi-Karma Economies\thanksref{footnoteinfo}} 

\thanks[footnoteinfo]{Research supported by NCCR Automation, a National Centre of
Competence in Research, funded by the Swiss National Science
Foundation (grant number $180545$).}

\author[ETH]{Ezzat Elokda} 
\author[ETH]{Andrea Censi} 
\author[ETH]{Saverio Bolognani}
\author[ETH]{Florian D\"orfler}
\author[ETH]{Emilio Frazzoli}

\address[ETH]{ETH Zurich, Zurich, 8092 Switzerland (e-mail: \{elokdae,acensi,bsaverio,dorfler,efrazzoli\}@ethz.ch).}

\begin{abstract}                
The large-scale allocation of public resources (e.g., transportation, energy) is among the core challenges of future \gls{CPHSs}.
In order to guarantee that these systems are efficient \emph{and} fair, recent works have investigated non-monetary resource allocation schemes, including schemes that employ \emph{karma}.
Karma is a non-tradable token that flows from users gaining resources to users yielding resources.
Thus far karma-based solutions considered the allocation of a single public resource, however, modern \gls{CPHSs} are complex as they involve the allocation of multiple coupled resources.
For example, a user might want to trade-off fast travel on highways for convenient parking in the city center, and different users could have heterogeneous preferences for such coupled resources.
In this paper, we explore how to optimally combine multiple \emph{karma economies} for coupled resource allocations, using two mechanism-design instruments: (non-uniform) karma redistribution; and (non-unit) exchange rates.
We first extend the existing \gls{DPG} model that predicts the \gls{SNE} of the multi-karma economies.
Then, in a numerical case study, we demonstrate that the design of redistribution significantly affects the coupled resource allocations, while non-unit exchange rates play a minor role.
To assess the allocation outcomes under user heterogeneity, we adopt \emph{Nash welfare} as our social welfare function, since it makes no interpersonal comparisons and it is axiomatically rooted in social choice theory.
Our findings suggest that the simplest mechanism design, that is, uniform redistribution with unit exchange rates, also attains maximum social welfare.
\end{abstract}


\end{frontmatter}

\input{sections/introduction}
\input{sections/model}
\input{sections/numerical}
\input{sections/conclusion}

\begin{ack}
We would like to thank Patrick Oberlin for the fruitful and interesting discussions.
\end{ack}

\bibliography{ifacconf}             
                                                   








\end{document}

%% file: sections/introduction.tex
\section{Introduction}

Control and automation will play a critical role in the large-scale allocation of public resources, including energy, transportation, and smart infrastructures, and must accomplish these societal-scale tasks efficiently and fairly~\citep{annaswamy2023control}.
Traditionally, \emph{\acrfull{CPHSs}} are studied in isolation, while public resources in modern mega-cities are complex and highly intertwined, presenting both opportunities and challenges.
The opportunities lie in added flexibility to devise social incentives, e.g., incentivizing \gls{EV} users to charge during peak traffic could reduce the traffic; and commuters could be willing to accept some travel delay in exchange for convenient parking.
The challenges lie in ensuring that such gains in \emph{efficiency} do not compromise \emph{fairness}, in consideration that the human users could have heterogeneous private preferences for the different resources.
Therefore, while incentivizing consumption of one public resource over another could be desirable overall, it could disfavor some groups.

In economics, the benefits of \emph{free trade} between different economies or resource domains are well studied~\citep{friedman2017price}, although this classical topic remains heavily debated until today due to concerns about externalities and inequities~\citep{antweiler2001free,irwin2020free}.
Specifically in engineering contexts, many works have considered the coupling of energy and transportation resources~\citep{alizadeh2016optimal,xu2018planning,cenedese2022incentive}.
For example, \cite{cenedese2022incentive} propose to de-congest highways by offering \gls{EV} users discounts on charging during peak traffic times.
However, these works rely on classical monetary instruments to facilitate the coupling of different resources, which has an unmodelled effect of favoring wealthier users and raises severe fairness concerns~\citep{arnott1994welfare,taylor2010addressing}.

This paper instead investigates how to couple different resources in a non-monetary manner, building on the recently proposed concept of \emph{karma economies}~\citep{elokda2023self,elokda2024carma}.
Karma is a scarce, non-tradable credit that flows from users consuming public resources to users yielding these resources.
Karma leverages the repetitive nature of many resource allocations to devise incentives in a \emph{self-contained} and thereby fair manner: over time, karma gives all users an equal opportunity to access the resource; meanwhile it is in the users' self-interest to prioritize access when they have highest \emph{urgency}.
Thus far previous works have focused on modelling \emph{single resource} karma economies as \emph{\gls{DPGs}}~\citep{elokda2024dynamic,elokda2023self,elokda2024carma} and other kinds of mean-field games~\citep{salazar2021urgency}.

Our paper's contributions are summarized as follows. First, in Section~\ref{sec:model}, we extend the previous karma \gls{DPG} model to incorporate resource-specific karma accounts for multiple resources.
The extended model features novel design elements
including (non-unit) karma exchange rates, non-uniform redistribution of karma, and state-dependent future discounting.
The existence of a \emph{\acrfull{SNE}} is guaranteed in the extended setting.
Second, in Section~\ref{sec:numerical}, we perform a real-world-inspired numerical case study of the coupled allocation of highway express lanes and priority parking spaces.
This study provides initial insights on how to optimally couple karma economies: it is found that combining economies leads to Pareto improvements over separate economies in most, but not all, considered karma designs.
Moreover, different designs prioritize different classes of users, and there is a trade-off to be made.
Of independent interest is the \emph{social welfare function} with which we evaluate these trade-offs: rather than summing the payoffs of heterogeneous users, which requires strong \emph{interpersonal comparability} assumptions, we adopt \emph{Nash welfare}, which is the unique social welfare function satisfying the classical axioms of social choice theory without making interpersonal comparisons~\citep{kaneko1979nash,roberts1980interpersonal}.
The paper is thus concluded in Section~\ref{sec:conclusion}.

\subsection{Notation}
Let $a,d \in D \subseteq \Nat$  and let $c \in C\subseteq \Real^n$, then
for a function $f : D \times C \rightarrow \Real$, we distinguish discrete and continuous arguments through the notation $f[d](c)$.
Similarly, $p[a \mid d](c)$ denotes the conditional probability of $a$ given $d$ and $c$. 
We use the shorthand notation $\sum_d f[d]$ (respectively, $\prod_d f[d]$) to denote $\sum_{d \in D} f[d]$ (respectively, $\prod_{d \in D} f[d]$).
We denote by $\delta \in \Delta(D):=\left\{\left. \sigma \in \Real_+^{|D|} \right\rvert \sum_d \sigma[d] = 1 \right\}$ a distribution over the elements of $D$, with $\delta[d] \in [0, 1]$ denoting the proportion of element $d$.
Finally, when considering heterogeneous user types, we denote by $x_\tau$ a quantity associated to type $\tau$.

%% file: sections/model.tex
\section{Multi-Karma Economy Model}
\label{sec:model}

Our game-theoretical model is adopted from~\cite{elokda2024carma} and extended to consider multiple resources and karma accounts.

\subsection{Karma-Based Multi-Resource Allocations}
\begin{figure}[t]
    \centering
    \includegraphics[width=0.48\textwidth]{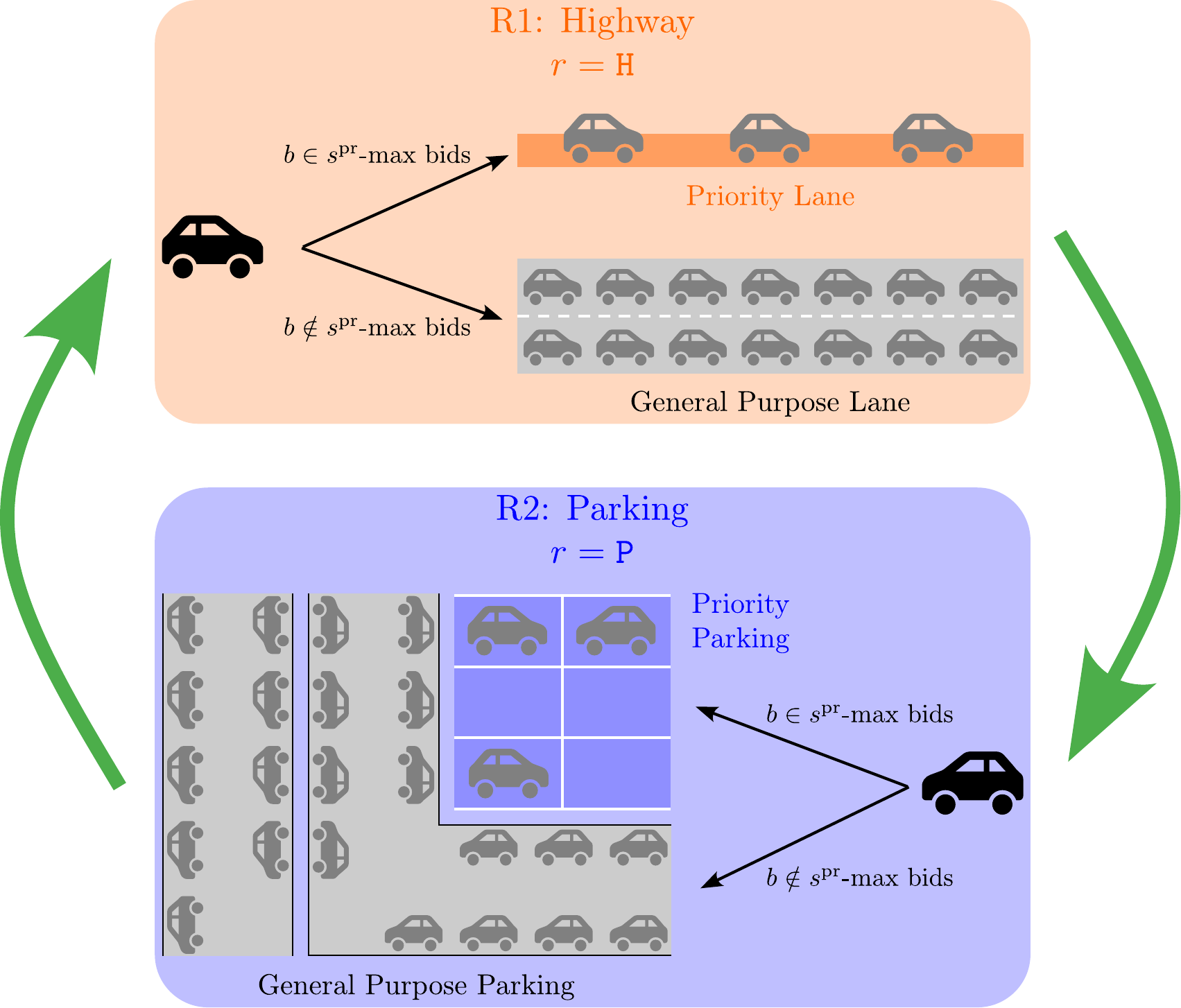}
    \caption{Example of coupled public resources.}
    \label{fig:multi-karma-economies}
\end{figure}

We consider an infinitely repeated setting over days.
On each day, a large population of users compete sequentially over $r \in \R = \{1,\dots,n_r\}$ public resources.
Fig.~\ref{fig:multi-karma-economies} illustrates a running example inspired by the daily commute to work.
In this example, users must first travel through a congested highway (denoted by $r=\highway$); then find parking in the congested city center (denoted by $r=\parking$).
Each resource $r$ has a total capacity $s[r] \in [0,1)$, expressed as a fraction of the total population, of which $\spr[r] \in [0,s[r])$ is dedicated for congestion-free, priority access (e.g., a highway express lane; or a reserved parking lot), and the remaining $\sgp[r] = s[r] - \spr[r]$ is left for potentially congested, general purpose access.
In particular, in case more than $\sgp[r]$ users are granted general purpose access, a time delay proportional to the excess demand is incurred by those users.

Priority access to each resource is regulated by means of a \emph{karma economy}.
Each user is endowed with a non-tradable, resource-specific karma credit $k[r] \in \K[r] = \{0,\dots,\kmax[r]\}$\footnote{$\kmax[r] \in \Nat$ ensures the state space is finite and simplifies the analysis; however, all our results can be extended to $\kmax[r]=\infty$ since our system is \emph{karma preserving}, cf.~\cite{elokda2023self}.}, and $K = \left[k[1],\dots,k[n_r]\right] \in \prod_r \K[r]$ denotes the vector of karma credits. The user may use its karma to place a bid $b \in \Nat$ for priority access.
The highest $\spr[r]$ bidders gain priority access and pay their bids, while all other bidders get general purpose access and do not make a payment.
The total payment is redistributed to the users according to a redistribution rule to be described hereafter.

\begin{figure}[t]
    \centering
    \begin{subfigure}[b]{0.4\textwidth}
        \centering
        \includegraphics[width=\textwidth]{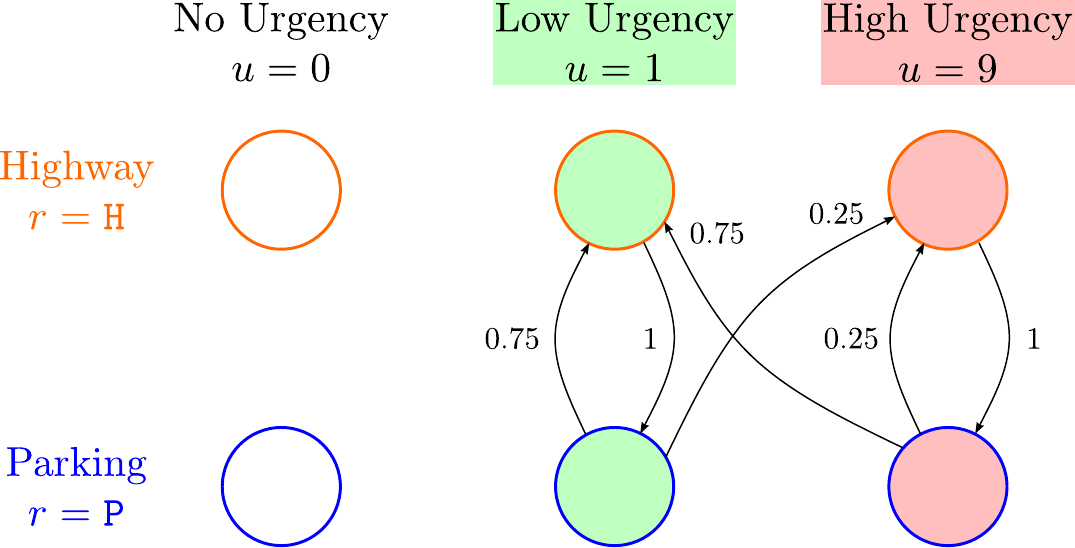}
        \caption{Suburb type $\suburb$.}
    \end{subfigure}

    \bigskip
    
    \begin{subfigure}[b]{0.4\textwidth}
        \centering
        \includegraphics[width=\textwidth]{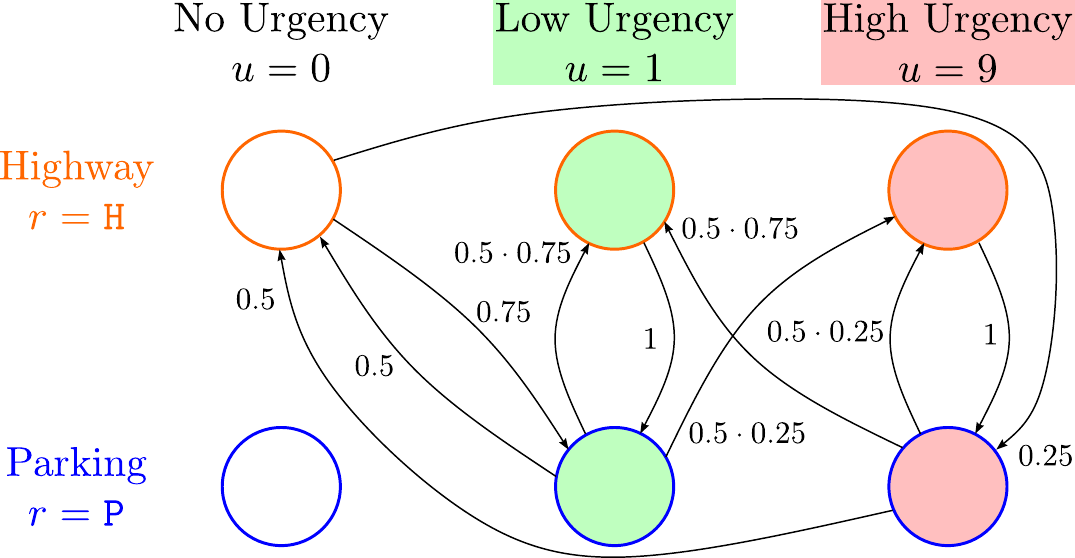}
        \caption{City type $\city$.}
    \end{subfigure}
    \caption{Example of resource-urgency Markov chain for two user types.
    Type $\suburb$ needs both resources every day, with urgency drawn independently at the beginning of the day ($\Prob[u=1]=0.75$, $\Prob[u=9]=0.25$).
    Type $\city$ is identical to $\suburb$, except that it needs resource $\highway$ only on half of the days.}
    \label{fig:Markov-chain}
\end{figure}

Not all users have the same \emph{urgency} to access the different resources.
We consider a finite number of user types $\tau \in \T = \{1, \dots, n_\tau\}$, with the type distribution denoted by $g \in \Delta(\T)$.
At the very beginning, each user is associated with an \emph{urgency state} $u \in \U = \{0, u_1,\dots,u_{n_u-1}\} \subset \Real_+$ for the first resource $r=1$.
Then, the joint resource-urgency state $[r,u]$ of each user evolves in time according to an exogenous, type-dependent Markov chain, denoted by $\phi_\tau[r^+,u^+ \mid r,u]$.
In order to model the sequential nature of the resource competitions, $\phi_\tau[r^+,u^+ \mid r,u]$ is required to satisfy, for all $\tau \in \T$, $u \in \U$,
\begin{subnumcases}{}
    \sum_{u^+} \phi[r+1,u^+ \mid r, u] = 1, & $r \in \{1,\dots,n_r - 1\}$, \label{eq:same-day-transition} \\
    \sum_{u^+} \phi[1,u^+ \mid r, u] = 1, & $r = n_r$, \label{eq:next-day-transition}
\end{subnumcases}
where~\eqref{eq:same-day-transition} captures transitions to the next resource competition on the same day, and~\eqref{eq:next-day-transition} captures transitions to the first resource competition on the next day.
Importantly, state $[r,u=0]$ models that the user \emph{does not require access} to resource $r$ on the current day.
In this state, the user will typically not actively compete for the resource, which we denote with the special bid symbol $b = \dnt$.
Fig.~\ref{fig:Markov-chain} illustrates a running example of $\phi_\tau[r^+,u^+ \mid r,u]$ for two types $\tau \in \T=\{\suburb,\city\}$: \emph{suburb type} $\suburb$ must use the highway and park every day; and \emph{city type} $\city$ needs the highway only occasionally, but must also park every day.

\subsection{Karma Design Instruments}
We explore the following (non-exhaustive) set of design methods to combine the different karma economies.

\subsubsection{Karma Redistribution.}
Karma payments for resource $r$ are redistributed to the karma accounts of $r$, according to one of the following redistribution schemes:
\begin{itemize}
    \item \emph{Redistribution to active:} All users that actively consumed the resource, i.e., bid $b \neq \dnt$, receive a uniform share of the redistribution.
    
    \item \emph{Redistribution to all:} All users receive a uniform share of the redistribution, including those that did not actively consume the resource ($b = \dnt$).
\end{itemize}

\subsubsection{Exchange Rates.}
The \emph{maximum bid} for resource $r$ is constrained by $\bmax[r,K] = \floor{\sum_{r'} \chi[r,r'] \: k[r']}$, where $\chi \in \Real_+^{n_r \times n_r}$ is a matrix of \emph{exchange rates}, with $\chi[r,r']$ denoting the rate at which karma of resource $r'$ can be used to place bids for resource $r$.
Naturally, we set $\chi[r,r]=1$, for all $r \in \R$,
and for two different resources $r \in \R$, $r' \ne r$, we will explore the following design regimes:
\begin{itemize}
    \item \emph{No Exchange:} $\chi[r,r'] = 0$;
    
    \item \emph{Unit Exchange:} $\chi[r,r'] = 1$; and
    
    \item \emph{Non-unit Exchange:} $\chi[r,r'] \neq 1$, $\chi[r,r'] \: \chi[r',r] = 1$.

\end{itemize}
Once a bid $b \leq \bmax[r,K]$ is placed,
payments are debited first from the current resource account $k[r]$ until depletion, then from $k[r+1]$ at rate $\frac{1}{\chi[r,r+1]}$, and so on.


\subsection{Individual Strategic Problem}
We now turn to model the coupled strategic problems of the individual users as a \emph{\acrfull{DPG}}~\citep{elokda2024dynamic}.
In addition to the static type $\tau \in \T$, each user is associated with the time-varying state $x=[r,\tilde x] \in \X = \R \times \tilde \X$, $\tilde x = [u,k] \in \tilde \X = \U \times \K$, where the discrete time-steps correspond to resource competition instances within days.
The state $x$ is composed of the \emph{public} component $r$ and the \emph{private} component $\tilde x = [u,k]$.
Conditional on the public $r$, the \emph{type-state distribution} is denoted by $d \in \D = \Delta(\T \times \tilde \X)^{n_r}$, with $d_\tau[\tilde x \mid r]$ giving the proportion of users in $[\tau,\tilde x]$ that compete over $r$.
In each time-step, users place a bid $b \in \B[r,K] = \{\dnt,0,\dots,\bmax[r,K]\}$ according to the \emph{policy} $\pi : \T \times \X \rightarrow \Delta(\B[r,K])$, with $\pi_\tau[b \mid x]$ denoting the probability of bidding $b$ when in $[\tau,x]$, and the space of policies is $\pi \in \Pi$.
The \emph{social state} is $(d,\pi) \in \D \times \Pi$, which gives the distribution of the private type-states $[\tau,\tilde x]$ and actions $b$.
Each individual user faces a \emph{\gls{MDP}} that is coupled to others through $(d,\pi)$.
In what follows, we specify the key elements of this \gls{MDP}: the \emph{immediate payoff function} $\sigma_\tau(d,\pi)$, and the \emph{state transition function} $p_\tau(d,\pi)$.

\subsubsection{Immediate Payoff Function $\sigma_\tau[x,b](d,\pi)$.}
Upon placing bid $b$, the user is assigned to an \emph{outcome} \mbox{$o \in \Out = \{\pr,\gp,\dnt\}$} according to a probability $\psi[o \mid r,b](d,\pi)$, and incurs a \emph{congestion delay} $\tqueue[r](d,\pi)$ in case $o = \gp$.
The immediate payoff is thus given by
\begin{multline}
    \label{eq:rewards}
    \sigma_\tau[r,u,b](d,\pi) \\
    \small
    = \begin{cases}
        0, & b = \dnt, \\
        u \: \left(\sigmanom - \psi[\gp \mid r, b] \: \tqueue[r]\right), & b \neq \dnt, \; u > 0, \\
        - \psi[\gp \mid r, b] \: \tqueue[r], & b \neq \dnt, \; u = 0,
    \end{cases}
\end{multline}
where, here and in what follows, we omit the dependency on $(d,\pi)$ on the right-hand side for notational convenience.
In case the user does not actively consume the resource ($b = \dnt$), it receives no payoff.
In case the user actively consumes the resource because it must ($b \neq \dnt$, $u > 0$), it receives a \emph{nominal payoff} $\sigmanom > \max_{r,(d,\pi)} t^d[r](d,\pi)$, minus the expected delay $\psi[\gp \mid r, b] \: \tqueue[r]$, weighed by the urgency $u$.
Finally, a user could actively consume the resource despite of no need ($b \neq \dnt$, $u = 0$), in which case it incurs the expected delay cost without gaining any payoff.

The derivation of $\psi[o \mid r,b](d,\pi)$ and $\tqueue[r](d,\pi)$ follows closely from~\cite{elokda2024carma}, and we give the final expressions only for brevity.
Let $\nu[b \mid r](d,\pi)$ be the proportion of users bidding $b$ for resource $r$, i.e., 
\begin{align}
    \nu[b \mid r](d,\pi) = \sum_{\tau,\tilde x} d_\tau[\tilde x \mid r] \: \pi_\tau[b \mid r,\tilde x].
\end{align}
Then, the probability of outcome $o=\pr$ is given by 
\begin{multline}
    \label{eq:prob-outcome-cont}
    \psi[\pr \mid r,b](d,\pi) \\
    \small = \begin{cases}
    1, & \sum_{b'> b} \nu[b' \mid r] \leq \spr[r] - \epsilon -  \nu[b \mid r], \\
    0, & \sum_{b'> b} \nu[b' \mid r] \geq \spr[r], \\
    \frac{\spr[r]-\sum_{b'>b}\nu[b' \mid r]}{\epsilon + \nu[b \mid r]}, & \text{otherwise},
    \end{cases}
\end{multline}
and it holds that $\psi[\dnt \mid r, b=\dnt] = 1$, $\psi[\dnt \mid r, b\ne\dnt] = 0$, and $\psi[\gp \mid r,b](d,\pi) = 1 - \psi[\pr \mid r,b] - \psi[\dnt \mid r, b]$.
The congestion delay is given by
\begin{align}
    \tqueue[r](d,\pi) = \max \left\{\frac{\sum_{b} \nu[b \mid r] \: \psi[\gp \mid r,b] - \sgp}{\sgp},0\right\}.
\end{align}

Notice that in~\eqref{eq:prob-outcome-cont}, the priority access capacity $\spr[r]$ is slightly under-allocated by the small parameter $\epsilon$, which is needed to ensure that $\psi[\pr \mid r,b](d,\pi)$ is continuous in $(d,\pi)$~\citep{elokda2024carma}.

\subsubsection{State transition function $p_\tau[x^+ \mid x, b](d,\pi)$.}
Given bid $b$, the state of the user evolves according to
\begin{multline}
    p_\tau[r^+,u^+,K^+ \mid r,u,K,b](d,\pi) \\ = \phi_\tau[r^+,u^+ \mid r,u] \sum_o \psi[o \mid r, b] \: \kappa[K^+ \mid r,K,b,o],
\end{multline}
where $\phi_\tau[r^+,u^+ \mid r,u]$ is the exogenous resource-urgency Markov chain, and $\kappa[K^+ \mid r,K,b,o](d,\pi)$ is the \emph{karma transition function} that encodes the karma payment and redistribution rule.
Let $\hat K$ be the user's karma after payment but before redistribution.
If there are only two resources, its transition probabilities are given by
\begin{multline}
    \label{eq:transition-payment}
    \Prob[\hat K \mid r, K, b, o] \\
    \small
    = \begin{cases}
        1, & o \neq \pr, \; \hat K = K, \\
        1, & o = \pr, \; b \leq k[r], \; \hat k[r] = k[r] - b, \; \hat k[r'] = k[r'], \\
        1, & o = \pr, \; b > k[r], \; \hat k[r] = 0, \hat k[r'] = k[r'] - \frac{b - k[r]}{\chi[r,r']}, \\
        0, & \text{otherwise}.
    \end{cases}
\end{multline}
where $r'\neq r$ is the resource not currently contested.
If there are more than two resources, Equation~\eqref{eq:transition-payment} can be readily extended by analogously incorporating more cases.

The \emph{average payment} to redistribute is given by
\begin{align}
    \paybar[r](d,\pi) = \sum_{b \neq \dnt} \nu[b \mid r] \: \psi[\pr \mid r,b] \: b.
\end{align}

In case of \emph{redistribution to all}, all users gain $\paybar[r](d,\pi)$, i.e.,
\begin{multline}
\label{eq:tansition-redist-all}
    \Prob^\all[K^+ \mid r, \hat K, o](d,\pi) \\
    \small 
    = \begin{cases}
        1, & k^+[r] = \hat k[r] + \paybar[r], \; k^+[r' \neq r] = \hat k[r'], \\
        0, & \text{otherwise}.
    \end{cases}
\end{multline}
Instead, in case of \emph{redistribution to active}, the \mbox{$\nu[\dnt \mid r](d,\pi)$} non-active users do not receive redistribution, while all other users gain $\paytilde[r](d,\pi) = \paybar[r] / (1 - \nu[\dnt \mid r])$, i.e.,
\begin{multline}
\label{eq:tansition-redist-active}
    \Prob^\act[K^+ \mid r, \hat K, o](d,\pi) \\
    \small
    = \begin{cases}
        1, & o = \dnt, \; K^+ = \hat K, \\
        1, & o \neq \dnt, \; k^+[r] = \hat k[r] + \paytilde[r], \; k^+[r' \neq r] = \hat k[r'], \\
        0, & \text{otherwise}.
    \end{cases}
\end{multline}

Putting everything together, we get
\begin{multline}
    \kappa[K^+ \mid r,K,b,o](d,\pi) \\
    = \sum_{\hat K} \Prob[K^+ \mid r,\hat K,o] \: \Prob[\hat K \mid r,K,b,o].
\end{multline}

Notice that two simplifications are made in Equations~\eqref{eq:transition-payment}, \eqref{eq:tansition-redist-all}--\eqref{eq:tansition-redist-active} for brevity.
First, it is supposed that $\paybar[r]$ is integer-valued (and similarly for $\paytilde[r]$ and $\frac{b-k[r]}{\chi[r,r']}$).
This can be readily extended by redistributing $\ceil{\paybar[r]}$ and $\floor{\paybar[r]}$ probabilistically such that $\paybar[r]$ is gained in expectation, cf.~\cite{elokda2023self,elokda2024carma}.
Second, it is supposed in~\eqref{eq:tansition-redist-all}--\eqref{eq:tansition-redist-active} that $k^+[r] \leq \kmax[r]$.
Instead, one could limit the redistribution to the users who will exceed $\kmax[r]$, and appropriately adjust $\paybar[r]$ (respectively, $\paytilde[r]$) to redisrtibute more karma to the others.

\subsection{Existence of \acrfull{SNE}}

Given the constituents of the coupled \gls{MDP}s $\sigma_\tau(d,\pi)$ and $p_\tau(d,\pi)$, we define the expected immediate rewards $R_\tau(d,\pi)$, state transition matrix $P_\tau(d,\pi)$, infinite horizon value function $V_\tau(d,\pi)$, and state-action value function $Q_\tau(d,\pi)$ respectively as
\begin{subequations}
\begin{align*}
    &R_\tau[r,u,K] = \sum_b \pi_\tau[b \mid r,u,K] \: \zeta_\tau[r,u,b], \\
    &P_\tau[x^+ \mid x] = \sum_b \pi_\tau[b \mid x] \: p_\tau[x^+ \mid x, b], \\
    &V_\tau[r,u,K] = R_\tau[r,u,K] + \alpha[r] \sum_{x^+} P_\tau[x^+ \mid r,u,K] \: V_\tau[x^+], \\
    &Q_\tau[r,u,K,b] = \zeta_\tau[r,u,b] + \alpha[r] \sum_{x^+} p_\tau[x^+ \mid r,u,k,b] \: V_\tau[x^+],
\end{align*}
\end{subequations}
where $\alpha[r]$ is a \emph{state-dependent discount factor} satisfying $\alpha[r \neq n_r] = 1$ and $\alpha[r = n_r] \in [0, 1)$, i.e., discounting occurs only on transitions to future days, but not within the same day.
Notice that $\prod_r \alpha[r] < 1$, and therefore the Bellman recursion $V_\tau$ is \emph{eventually contracting} and has a unique solution that is continuous in $(d,\pi)$~\citep{stachurski2021dynamic}.

\begin{definition}
A social state $(d^*,\pi^*)$ is a \emph{\acrfull{SNE}} if, for all $\tau \in \T$, $x = [r,\tilde x] \in \X$,
\begin{subequations}
\begin{align}
    d^*_\tau[\tilde x \mid r] &= \sum_{\tilde x'} d^*_\tau[\tilde x' \mid r] \: P_\tau[r,\tilde x \mid r-1,\tilde x'](d^*,\pi^*), \label{eq:SNE-1} \\
    \pi^*_\tau[\cdot \mid x] &\in \argmax_{\delta \in \Delta(\B[x])} \sum_b \delta[b] \: Q_\tau[x,b](d^*,\pi^*), \label{eq:SNE-2} 
\end{align}
\end{subequations}
where for $r=1$ we define $r-1 := n_r$.
\end{definition}


The existence of a \gls{SNE} is guaranteed for general \gls{DPG}s if the immediate payoff function $\sigma_\tau(d,\pi)$ and the state transition function $p_\tau(d,\pi)$ are continuous $(d,\pi)$~\citep{elokda2024dynamic}.
The following theorem is thus immediate from~\cite[Proposition~1]{elokda2024dynamic}.

\begin{theorem}
A \gls{SNE} $(d^*,\pi^*)$ is guaranteed to exist in the multi-karma economy.
\end{theorem}

In what follows, we utilize the \emph{evolutionary dynamics}-inspired algorithm developed in~\cite{elokda2024dynamic,elokda2023self} to investigate the \gls{SNE} numerically.

%% file: sections/numerical.tex
\section{Numerical Analysis}
\label{sec:numerical}

\subsection{Individual and Social Welfare Measures}

We define the \emph{individual welfare} of a user of type $\tau$ at a \gls{SNE} $(d^*,\pi^*)$ as the \emph{long-run expected average payoff}, denoted by $\sigmabar_\tau(d^*,\pi^*)$.
We consider two definitions of $\sigmabar_\tau(d^*,\pi^*)$ which differ in whether time-steps in which the user does not actively consume the resource (i.e., bids $b = \dnt$) are treated as \emph{endogenous} or \emph{exogenous} and included in the average payoff or not, given by
{\small
\begin{align}
    \sigmabarendo_\tau(d^*,\pi^*) &= \frac{\sum_{r,u,K,b} d^*_\tau[u,K \mid r] \: \pi^*_\tau[b \mid r,u,K] \: \sigma_\tau[r,u,b]}{\sum_{r,u,K,b} d^*_\tau[u,K \mid r] \: \pi^*_\tau[b \mid r,u,K]}, \label{eq:average-payoff-endo} \\
    \sigmabarexo_\tau(d^*,\pi^*) &= \frac{\sum_{r,u,K,b \neq \dnt} d^*_\tau[u,K \mid r] \: \pi^*_\tau[b \mid r,u,K] \: \sigma_\tau[r,u,b]}{\sum_{r,u,K,b \neq \dnt} d^*_\tau[u,K \mid r] \: \pi^*_\tau[b \mid r,u,K]}. \label{eq:average-payoff-exo}
\end{align}
}

To assess \emph{social welfare}, we adopt the \emph{Nash welfare function}, given by
\begin{align}
    \sw(d^*,\pi^*) = \sum_\tau g_\tau \: \log(\sigmabar_\tau - \sigmabarbench_\tau), \label{eq:SW-log}
\end{align}
where $\sigmabarbench_\tau$ is the \emph{long-run average payoff in a benchmark or status-quo allocation that makes all users simultaneously worst off}.
In comparison to utilitarianism (sum of payoffs), egalitarianism (minimum payoff), and other common social welfare measures, Nash welfare is the \emph{unique}\footnote{That is, unique up to monotonic transformation.} social welfare function satisfying the classical axioms of social choice theory with \emph{no interpersonal comparisons}~\citep{kaneko1979nash,roberts1980interpersonal}.
Intuitively, maximizing~\eqref{eq:SW-log} corresponds to maximizing \emph{relative improvements to the benchmark} with no regard to the magnitude or scale of different users' payoffs, since the $\log$ is invariant to scale.
This property is especially important in our setting which does not have an inter-personally comparable measure of payoff (typically assumed to be money).

As the benchmark allocation in~\eqref{eq:SW-log}, we naturally consider the allocation in which resource access is left uncontrolled (i.e., $\spr[r]=0$, for all $r \in \R$), leading to congestion delays that are equally endured by the active users regardless of urgency.
Notice that $\sigmabarbench_\tau$ can be derived analogously to $\sigmabar_\tau$ with $\spr[r]=0$.

\subsection{Results and Discussion}

\begin{table*}[t]
\caption{Individual and social welfare measures under different multi-karma designs. For each measure, the maximum (respectively, minimum) attained is marked green (respectively, red).}
\centering
\label{tab:results}
\begin{tabular}{ccr||c|c|c|c|c|c}
\toprule
& & & \multicolumn{6}{c}{\textbf{Individual and Social Welfare}} \\
& & & \multicolumn{3}{c|}{Endogenous Inactivity} & \multicolumn{3}{c}{Exogenous Inactivity} \\
& & & \begin{tabular}{@{}c@{}} Type $\suburb$ \\ $\sigmabarendo_\suburb$ \end{tabular} & \begin{tabular}{@{}c@{}} Type $\city$ \\ $\sigmabarendo_\city$ \end{tabular} & \begin{tabular}{@{}c@{}} Social Welfare \\ $\swendo$ \end{tabular} & \begin{tabular}{@{}c@{}} Type $\suburb$ \\ $\sigmabarexo_\suburb$ \end{tabular} & \begin{tabular}{@{}c@{}} Type $\city$ \\ $\sigmabarexo_\city$ \end{tabular} & \begin{tabular}{@{}c@{}} Social Welfare \\ $\swexo$ \end{tabular} \\
\hline \hline
\multicolumn{3}{r||}{Benchmark} & $3.7500$ & $2.6250$ & -- & $3.7500$ & $3.5000$ & -- \\
\hline \hline
\multirow{14}{*}[22pt]{\rotatebox[origin=c]{90}{\textbf{Karma Design}}} & \multirow{7}{*}[12pt]{\rotatebox[origin=c]{90}{\begin{tabular}{@{}c@{}}Redist. \\ to Active\end{tabular}}} & & & & & & \\[-7pt]
& & No Exchange & $4.5741$ & $3.2801$ & $-0.3082$ & $4.5741$ & $4.2658$ & $-0.2301$ \\
& & Unit Exchange &  $4.5649$ & $3.2750$ & $-0.3177$ & $4.5649$ & $4.1424$ & $-0.3236$ \\
& & Exchange $\parking > \highway$ & \cellcolor{green!25}$4.6178$ & $3.3018$ & $-0.2661$ & \cellcolor{green!25}$4.6178$ & $4.3071$ & $-0.1780$ \\
& & Exchange $\parking < \highway$ &  $4.4940$ & \cellcolor{red!25}$3.2618$ & \cellcolor{red!25}$-0.3735$ & $4.4940$ & \cellcolor{red!25}$4.0326$ & \cellcolor{red!25}$-0.4629$ \\
\cmidrule{2-9}
& \multirow{7}{*}[15pt]{\rotatebox[origin=c]{90}{\begin{tabular}{@{}c@{}}Redist. \\ to All\end{tabular}}} & & & & & & \\[-10pt]
& & No Exchange & $4.4749$ & $3.3300$ & $-0.3356$ & $4.4749$ & $4.4401$ & $-0.1918$ \\
& & Unit Exchange &  $4.5020$ & \cellcolor{green!25}$3.4436$ & \cellcolor{green!25}$-0.2425$ & $4.5020$ & \cellcolor{green!25}$4.5915$ & \cellcolor{green!25}$-0.0987$ \\
& & Exchange $\parking > \highway$ & $4.5238$ & $3.4065$ & $-0.2515$ & $4.5238$ & $4.5420$ & $-0.1076$ \\
& & Exchange $\parking < \highway$ & \cellcolor{red!25}$4.4356$ & $3.3192$ & $-0.3712$ & \cellcolor{red!25}$4.4356$ & $4.4257$ & $-0.2274$ \\
\bottomrule
\end{tabular}
\end{table*}

Table~\ref{tab:results} provides an overview of our findings.
We computed the \gls{SNE} for each of the karma design combinations shown in the table, with the following illustrative parameter settings: $s[\highway]=s[\parking] = 0.5$, $\spr[\highway]=0.1875$, $\spr[\parking]=0.2$, $g_\suburb = g_\city = 0.5$, $\kbar[\highway] = \kbar[\parking] = 8$, $\kmax[\highway] = \kmax[\parking] = 24$, $\sigmanom = 2$, $\alpha[\highway] = 1$, $\alpha[\parking] = 0.98$, $\epsilon = 10^{-4}$, and the resource-urgency Markov chain shown in Fig.~\ref{fig:Markov-chain}.
In \emph{exchange $\parking > \highway$} (respectively, \emph{exchange $\parking < \highway$}), we set $\chi[\highway,\parking] = 3/2$ and $\chi[\parking,\highway] = 2/3$ (respectively, $\chi[\highway,\parking] = 2/3$ and $\chi[\parking,\highway] = 3/2$), such that one unit of $\parking$-karma exchanges for more (respectively, less) $\highway$-karma.
The main insights from Table~\ref{tab:results} are summarized as follows.
Overall, there are significant Pareto improvements with respect to the benchmark in all welfare measures under all karma designs (endogenous type $\suburb$: $18.3$--$23.1\%$; endogenous type $\city$: $24.3$--$31.2\%$; exogenous type $\suburb$: $18.3$--$23.1\%$; exogenous type $\city$: $15.2$--$31.2\%$).
Treating inactivity as endogenous or exogenous leads to qualitatively similar results, which suggests that these definitions can be used interchangeably.
In what follows we summarize our findings in regards to \emph{redistribution} and \emph{exchange rates}.

\subsubsection{Redistribution.} \emph{Redistribution to active} prioritizes type $\suburb$ users, while \emph{redistribution to all} prioritizes type $\city$ users; thus there is a trade-off between these two designs.
Under \emph{redistribution to active}, the type $\suburb$ users who are always active benefit from not redistributing karma to the occasionally inactive type $\city$ users.
However, according to the Nash welfare measure, which is maximized under \emph{redistribution to all} for almost all exchange rates, the relative improvement of type $\city$ under \emph{redistribution to all} outweighs the relative improvement of type $\suburb$ under \emph{redistribution to active}.
Intuitively, conditioned on $u>0$, type $\city$ faces greater congestion on average than type $\suburb$ (for $\city$, $66.7\%$ of time-steps with $u>0$ are in the more congested resource $\parking$; for $\suburb$, this fraction is only $50\%$). Therefore, type $\city$ has relatively more to lose by not receiving redistribution when inactive than type $\suburb$ has to gain.

\subsubsection{Exchange Rates.} 
\begin{figure}[t]
    \centering
    \begin{subfigure}[b]{0.24\textwidth}
        \centering
        \includegraphics[width=\textwidth]{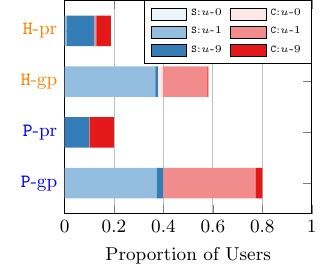}
        \caption{Redist. Active, No Exchange}
        \label{fig:active-no-exchange}
    \end{subfigure}
    \hfil
    \begin{subfigure}[b]{0.24\textwidth}
        \centering
        \includegraphics[width=\textwidth]{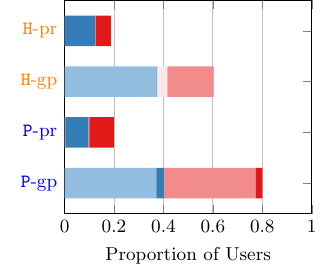}
        \caption{Redist. Active, Unit Exchange}
        \label{fig:active-full-exchange}
    \end{subfigure}

    \bigskip

    \begin{subfigure}[b]{0.24\textwidth}
        \centering
        \includegraphics[width=\textwidth]{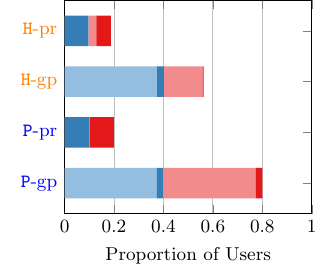}
        \caption{Redist. All, No Exchange}
        \label{fig:all-no-exchange}
    \end{subfigure}
    \hfil
    \begin{subfigure}[b]{0.24\textwidth}
        \centering
        \includegraphics[width=\textwidth]{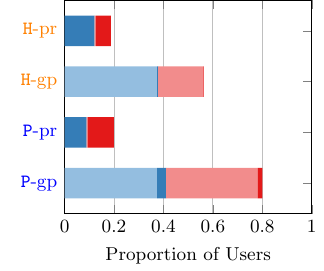}
        \caption{Redist. All, Unit Exchange}
        \label{fig:all-full-exchange}
    \end{subfigure}
    
    \caption{Resource utilization under different multi-karma designs.}
    \label{fig:resource-utilization}
\end{figure}

\emph{Unit exchange} achieves greater individual and social welfare than \emph{no exchange} under \emph{redistribution to all}, but, rather surprisingly, \emph{not} under \emph{redistribution to active}.
Fig.~\ref{fig:resource-utilization}, which shows the resource utilization per user type and urgency, provides insight on this finding.
An important observation of Fig.~\ref{fig:active-no-exchange}--\ref{fig:active-full-exchange} is that under \emph{redistribution to active}, \emph{unit exchange} incentivizes more of the type $\city$, $u=0$ users to consume resource $\highway$ despite having no need (cf. utilization of $\highway$-gp in Fig.~\ref{fig:active-full-exchange}).
Namely, the extra $\highway$-karma gained by those users is useful for the subsequent resource $\parking$.
This leads to increased congestion in $\highway$, and the decrease in individual and social welfare reported in Table~\ref{tab:results}.

In contrast, Fig.~\ref{fig:all-no-exchange}--\ref{fig:all-full-exchange} demonstrates how \emph{unit exchange} works out to everyone's benefit under \emph{redistribution to all}, under which $u=0$ users naturally have no incentive to consume resources.
Under \emph{no exchange} (Fig.~\ref{fig:all-no-exchange}), the isolated $\highway$-karma economy guarantees that priority access is split evenly between both user types, leading to some lowly urgent $\city$ users to occupy $\highway$-pr.
Under \emph{unit exchange} (Fig.~\ref{fig:all-full-exchange}), the type $\city$ users effectively exchange their low urgency share in $\highway$-pr for a larger high urgency share in $\parking$-pr.
This also benefits the type $\suburb$ users who subsequently always gain priority access to $\highway$ when highly urgent.

Finally, in regards to \emph{non-unit exchange rates}, Table~\ref{tab:results} suggests that overall, they do not provide pronounced benefits over \emph{unit exchange}, and in particular \emph{exchnge $\parking < \highway$} leads to worse welfare than \emph{no exchange}.
The non-unit exchange rates lead to \emph{inflation} of the karma available for one resource and \emph{deflation} for the other, which the users counteract by appropriately increasing/decreasing their bids.
We conjecture that the observed numerical deviations to \emph{unit exchange} are due to the quantization of the integer karma and that the karma saturation at $\kmax[r]$ is more or less binding at different exchange rates.

%% file: sections/conclusion.tex
\section{Conclusion}
\label{sec:conclusion}

We extended the karma \acrfull{DPG} model to enable the coupling of different resource allocations and explore a multitude of designs to achieve this coupling.
A \acrfull{SNE} is guaranteed to exist in the extended model.
Numerical analysis of the \gls{SNE} suggests that the simplest design is also the most robust and amenable to welfare improvements, that is, redistribute karma to all users (including inactive users), and allow exchange of karma between resources at unit rate.